\documentclass{amsart}

\usepackage{amsmath}
\usepackage{amsrefs}
\usepackage{amssymb}
\usepackage{graphicx}
\usepackage{comment}

\newcommand{\x}{\mathbf{x}}

\begin{document}

\title[Entropy measures as geometrical tools in the study of cosmology]{Entropy measures as geometrical tools\\ in the 
study of cosmology}

\author{G.~Weinstein}
\email{gilbertw@ariel.ac.il}
\address{Physics Department, Ariel University, Ariel 40700, Israel}
\address{Department of Mathematics, Ariel University, Ariel 40700, Israel}

\author{Y.~Strauss}
\email{yosefst@ariel.ac.il}
\address{Department of Mathematics, Ben Gurion University, Be'er Sheva 84105, Israel}

\author{S.~Bondarenko}
\email{sergeyb@ariel.ac.il}
\address{Physics Department, Ariel University, Ariel 40700, Israel}

\author{A.~Yahalom}
\email{asya@ariel.ac.il}
\address{Department of Electrical and Electronic Engineering, Ariel University, Ariel 40700, Israel}

\author{M.~Lewkowicz}
\email{lewkow@ariel.ac.il}
\address{Physics Department, Ariel University, Ariel 40700, Israel}

\author{L.~P.~Horwitz}
\email{larry@post.tau.ac.il}
\address{School of Physics and Astronomy, Raymond and Beverly Sackler Faculty of Exact Sciences,
Tel Aviv University, Tel Aviv 69978, Israel}
\address{Department of Physics, Bar Ilan University, Ramat Gan 52900, Israel}
\address{Physics Department, Ariel University, Ariel 40700, Israel}

\author{J.~Levitan}
\email{levitan@ariel.ac.il}
\address{Physics Department, Ariel University, Ariel 40700, Israel}

\begin{abstract}
Classical chaos is often characterized as exponential divergence of nearby trajectories.  In many interesting cases 
these trajectories can be identified with geodesic curves. We define here the entropy by $S = \ln \chi (x)$ with 
$\chi(x)$
being the distance between two nearby geodesics. We derive an equation for the entropy which by 
transformation to a Riccati    -type equation becomes similar to the Jacobi equation. We further 
show that the geodesic equation for a null geodesic in a double warped space time leads to the same 
entropy equation. By applying a Robertson-Walker metric for a flat three-dimensional Euclidian 
space expanding as a function of time, we again reach the entropy equation stressing the connection 
between the chosen entropy measure and time. We finally turn to the Raychaudhuri equation for 
expansion, which also is a Riccati     equation similar to the transformed entropy equation. Those Riccati    -
type equations have solutions of the same form as the Jacobi equation. The Raychaudhuri equation can 
be transformed to a harmonic oscillator equation, and it has been shown that the geodesic deviation 
equation of Jacobi is essentially equivalent to that of a harmonic oscillator. The Raychaudhuri equations 
are strong geometrical tools in the study of General Relativity and Cosmology. We suggest a refined 
entropy measure applicable in Cosmology and defined by the average deviation of the geodesics in a 
congruence.
\end{abstract}


\maketitle

\section{Introduction}

Classical chaos is generally defined as exponential divergence of nearby trajectories causing instability of the orbits 
with respect to initial conditions or quite simply as high sensitivity to initial conditions. The extent of divergence 
is quantified in terms of Lyapunov exponents measuring the mean rate of exponential separation of neighboring 
trajectories.

The norm  (for the Euclidean case, for example)
\[
 d(\tau) = \sqrt{\sum_{i=1}^n \delta x_i^2(\tau)}
\]
is a measure of the divergence of two neighboring trajectories, where $\delta x_i$ is the $i$-{th} component of the 
displacement between two nearby trajectories at time $t$.
The mean rate of exponential divergence is given by~\cite{tabor1989chaos} 
\[
 \omega=\lim_{\substack{\tau\to\infty\\d(0)\to0}} \frac1\tau \ln \left(\frac{d(\tau)}{d(0)}\right).
\]

The Kolmogorov  entropy is related to the Lyapunov exponents. It gives a measure of the amount of information lost or 
gained by the system as it evolves~\cite{tabor1989chaos}. It can be computed from the Lyapunov exponent by 
\[
  h_K = \int_P \sum_{\omega_i>0} \omega_i\, d\mu,
\]
which 
is the sum of all positive Lyapunov exponents averaged over some region of the phase space $P$ with measure          
$d\mu$.

One would naturally be interested in defining a measure for stochasticity in regions with divergence.
The function $d(\tau)$ initially has an irregular behavior, and evolves into a form in which the limit as $\tau\to\infty$ of
\[
 \frac1\tau\left(\frac{d(\tau)}{d(0)}\right)
\]
converges to a value that depends on the initial conditions. Casartelli et al~\cite{casartelli1976} argued that this 
quantity is deeply related to the Kolmogorov entropy and also exhibits 
strong sensitivity to the initial conditions.

Benettin et al defined a similar entropy~\cite{benettin1976} and calculated a Kolmogorov-like entropy for the 
Henon-Heiles system. 
However we shall take a different route in this study.

There are many interesting physically relevant  examples for which the trajectories can be put into correspondence with 
geodesic curves, for example, in problems in general relativity, and in the conformal map of Hamiltonian potential 
models~\cite{horwitz2007}, with geodesic deviation described in terms of a Jacobi equation related to the 
curvature.  In the following, we provide a relation between the Jacobi equation, the entropy (as defined above) and the geodesic equation itself.

Let there be given two nearby geodesics, $m$ and $n$, and let $\tau$ be the affine parameter on the geodesics. For
a point $\x$ with parameter $\tau$ on the geodesic $m$, one may define the geodesic deviation as the length of the 
shortest path from $m$ to $n$. Let's denote this geodesic deviation by $\chi(\tau)$.  The Jacobi equation states 
that~\cite{gutzwiller2014chaos}
\begin{equation} \label{jacobi}
 \frac{d^2\chi(\tau)}{d\tau^2} = - K(\x(\tau))\chi(\tau)
\end{equation}
where $K$ is the Gaussian curvature, and this simple form for the curvature is restricted to two-dimensional systems.

We employ the entropy defined by~\cite{Evangelidis1983}

\begin{equation} \label{entropy}
  S = \ln\chi(\tau)
\end{equation}
From Equations~\eqref{jacobi}-\eqref{entropy} we derive
\begin{equation} \label{S}
 \ddot S(\tau) + \left(\dot S(\tau)\right)^2 + K(\x(\tau)) = 0,
\end{equation}
where $\dot {}=d/d\tau$.
One may transform Equation~\eqref{S} to a Riccati    -type equation by letting $\dot S(\tau)=X(\tau)$
\begin{equation} \label{ricatti-K}
 \dot X(\tau) + X(\tau)^2 +K(\x(\tau))=0.
\end{equation}
The general Riccati     equation has the form
\begin{equation} \label{ricatti}
 \dot X(\tau) = q_0(\tau) + q_1(\tau) X + q_2(\tau) X^2.
\end{equation}
The solution to equation~\eqref{ricatti}  is  $X=-\dot u/q_2 u$  with $u$ being the solution to the equation
\begin{equation} \label{linear}
 \ddot u - T\dot u + Ru =0
\end{equation}
In Equation~\eqref{linear} $R=q_2q_0$  and  $T=q_1+\dot q_2/q_2$. We therefore 
obtain the  equation
\begin{equation} \label{curvature}
 \ddot u + Ku=0,
\end{equation}
This equation is the Jacobi equation in two dimensions.

For a flat space  $K=0$ and equation~\eqref{S} takes he form
\begin{equation} \label{K=0}
  \ddot S(\tau) + (S\dot (\tau))^2 =0,
\end{equation}
This equation has some resemblance with the geodesic equation
\begin{equation} \label{geodesic}
 \frac{d^2 x^\mu}{d\tau^2} + \Gamma^\mu_{\sigma\rho} \frac{dx^\sigma}{d\tau} \frac{dx^\rho}{d\tau} = 0
\end{equation}
in particular if $\Gamma^\mu_{\rho\sigma}$ vanishes for $\rho$ not equal to $\sigma$.

\section{Application to gravitation}

Consider, in particular, the geodesic equation for a null geodesic in a \emph{double-warped} spacetime 
\[
 ds^2 = -\phi^2dt^2 + a^2 g_{ij} dx^i dx^j
\]
where  $\phi=\phi(x)$, $a = a(t)$, and $g$ is independent of $t$.
Consider further a variation of the geodesic with $\delta x^i=0$
\[
 \delta s = \int \left(- \phi^2 \dot t \delta \dot t + aa' g_{ij} \dot x^i \dot x^j \delta t\right) d\tau = 0
\]
where ${}'=d/dt$.
Integrating by parts one gets $\int (\phi^2\ddot t + aa' g_{ij} \dot x^i \dot x^j)\delta t\, d\tau = 0$ which implies
\begin{equation} \label{geo}
 \phi^2\ddot t + aa' g_{ij} \dot x^i \dot x^j = 0,
\end{equation}
For a null geodesic one has 
\[
 \phi^2 \dot t^2 = a^2 g_{ij} \dot x^i \dot x^j.
\]
Substituting into the geodesic equation~\eqref{geo} above leads to
\begin{equation} \label{}
 \ddot t + \frac{a'}{a} \dot t^2 = 0.
\end{equation}
We have achieved the equation which formally is the same as the entropy equation~\eqref{K=0}.

Taking into account that the universe is evolving in time we study the entropy 
$S(\tau)=\ln\chi(\tau)$ in a four-dimensional cosmological spacetime with a time dependent metric. It is in fact a special 
case of a Robertson-Walker metric for a universe for which the space for a fixed time is a flat three-dimensional 
Euclidian space expanding as a function of time~\cite{carroll2004spacetime}.
\footnote{The model used here provides a simple illustration of the similarity between the geodesic equation and the entropy 
equation~\eqref{K=0}, which is the main intent of this study. Reference~\cite{carroll2004spacetime} also splits the geodesic equation into a time part and a space part but uses a different technique with the purpose of obtaining the
cosmological redshift.}
The metric of the model is given by 
\begin{equation} \label{metric}
 ds^2 = -dt^2 + a(t)^2 (dx^2+dy^2+dz^2).
\end{equation}
The Christoffel symbols for the time components $\mu=0$ are given by~\cite{carroll2004spacetime}
\[
 \Gamma^0_{00}=\Gamma^0_{i0}=\Gamma^0_{0i}=0, \quad 
 \Gamma^0_{ij} = a(t)\dot a(t) \delta_{ij}.
\]
By inserting these into the geodesic equation~\eqref{geodesic} we obtain
\begin{equation} \label{geo-time}
 \frac{d^2 x^0}{d\tau^2} + a(t)\dot a(t) \delta_{ij} \frac{dx^i}{d\tau} \frac{dx^j}{d\tau}=0.
\end{equation}
The Christoffel symbols for the spatial components ($\mu\ne0$) are
\[
 \Gamma^i_{jk} = \Gamma^i_{00}=0, \quad \Gamma^i_{j0}=\Gamma^i_{0j}=\frac{\dot a(t)}{a(t)} \delta^i_j.
\]
and the spatial part of the geodesic equation takes the form
\begin{equation} \label{spatial}
 \frac{d^2 x^i}{d\tau^2} + \frac{\dot a(t)}{a(t)} \delta^i_j \frac{dx^i}{d\tau} \frac{dx^j}{d\tau}=0.
\end{equation}
Equations~\eqref{geo-time} and~\eqref{spatial} constitute the splitting of the geodesic equation into the timelike and spacelike parts~\cite{carroll2004spacetime}.

For particles moving freely under purely gravitational forces one can find a freely falling coordinate system with the 
motion being a straight line in space time
\begin{equation} \label{freefall}
 \frac{d^2 x^a}{d\tau^2} = 0.
\end{equation}
Here $\tau$ is the proper time
\begin{equation} \label{time}
 d\tau^2 = \eta_{\alpha\beta}dx^\alpha dx^\beta
\end{equation}

For massless particles the RHS of equation~\eqref{time} vanishes~\cite{carroll2004spacetime}, and we may use 
$\sigma=x^0$ as the parameter instead of $\tau$. 
Photons follow null-geodesics and we restrict ourselves to paths along the x-axis,
i.e.\ $x^\mu(\sigma) = \{ t(\sigma), x(\sigma), 0, 0\}$. With the metric given by~\eqref{metric} and $ds^2=0$ we 
obtain
\begin{equation} \label{null}
 -dt^2 + a(t)^2 dx^2 =  0.
\end{equation}
This leads to the equation
\begin{equation} \label{null1}
 \frac{dx}{d\sigma} = \frac{1}{a(t)} \frac{dt}{d\sigma}
\end{equation}
By solving for $dt/d\sigma$ and inserting the null-condition~\eqref{null} into the time component
for the geodesic equation~\cite{Evangelidis1983} we finally achieve the equation
\begin{equation} \label{geo-null}
 \frac{d^2t}{d\sigma^2} + \frac{\dot a(t)}{a(t)} \left( \frac{dt}{d\sigma} \right)^2=0.
\end{equation}
This equation is formally identical to the entropy equation~\eqref{K=0}.

It is noteworthy that a resemblance between the geodesic equation and the entropy equation is obtained by inserting 
the null condition into the time part of the geodesic equation and not the spatial part, which underlines the connection 
between the present definition of entropy with time rather than space. 

\section{The Raychaudhuri equation}

The definition of entropy as defined by equation~\eqref{entropy} has its origin in the geodesic deviation equation 
describing the behavior of a one-parameter family of nearby geodesics and is, as remarked, in the present form 
restricted to systems of at most two dimensions. For higher dimensional systems one needs more refined tools to describe 
the behavior of a bundle of geodesics, the so-called congruence. We now argue that the Raychaudhuri equation may 
provide such tools in dimension $4$. In a forthcoming study we shall show examples of entropy defined by the average deviation of the geodesics in a congruence.

Let $\xi^i$ be the tangent vector field to a geodesic flow, and $h^{ij}$ be the metric on the subspace 
perpendicular to $\xi$. The Raychaudhuri equation is
\begin{equation} \label{raychaudhuri}
   \frac{d\theta}{d\tau} = -\frac13\theta^2 - \sigma_{ij}\sigma^{ij} + \omega_{ij}\omega^{ij} - R_{ij}\xi^i\xi_j,
\end{equation}
where $\tau$ is the affine parameter along the geodesic, and $R_{ab}$ is the Ricci tensor of the metric~\cite{kar2007}, 
$\theta=\nabla_i\xi_j  h^{ij}$  is the expansion, $\sigma_{ij}= \nabla_{(i}\xi_{j)}-\frac13\theta h_{ij}$ 
the shear, and $\omega_{ij}=\nabla_{[i}\xi_{j]}$ the twist.
Round brackets represent symmetrization, and square brackets represent anti-symmetrization.

For completeness, and because it is very simple, we carry out the derivation of this equation explicitly.
Denoting the covariant derivative $\nabla_j\xi_i$ by $\xi_{ij}$, the geodesic equation is:
\[
  \xi^j\xi_{ij}=0,
\]
and because $\xi^i\xi_i=\text{constant}$ it follows that also:
\[
  \xi^i\xi_{ij}=0.
\]
Without loss of generality we assume that $\xi^i\xi_i=-1$. The metric on the spacelike subspace perpendicular to $\xi$ 
is then
\[
  h_{ij} = g_{ij} + \xi_i\xi_j.
\]
We now decompose the derivative $\xi_{ij}$ of $\xi$ into three components:
\begin{align*}
  \theta &= h^{ij}\xi_{ij} = g^{ij} \xi_{ij}, \\[1ex]
  \sigma_{ij} &= \frac12 \left( \xi{ij}+\xi{ji} \right) - \frac13\theta h_{ij}, \\[1ex]
  \omega_{ij} &= \frac12 \left(\xi_{ij} - \xi_{ji} \right).
\end{align*}
We note that the expansion $\theta$ measures the logarithmic derivative of the 
volume element in the space perpendicular to $\xi$, the shear $\sigma_{ij}$ measures the
non-conformal part of the defomation of the metric $h$, and the twist $\omega_{ij}$ measures the 
entangling of the geodesic trajectories, i.e.\
the obstruction to $\xi$ being hypersurface-orthogonal. The expansion $\theta$ in~\eqref{raychaudhuri} corresponds to $\dot S$ in~\eqref{entropy} and can be taken as the derivative of the entropy. Equation~\eqref{entropy} is the $2$-dimensional version of~\eqref{raychaudhuri}.

We can decompose:
\[
  \xi_{ij} = \frac13 \theta h_{ij} + \sigma_{ij} + \omega_{ij},
\]
and note that these three components are mutually orthogonal:
\[
  \frac13 \theta h_{ij} \sigma^{ij} = \frac13\theta h_{ij} \omega^{ij} = \sigma_{ij} \omega^{ij} = 0.
\]
The first expression on the left vanishes because $\sigma$ is traceless, the second and the third vanish because
$h_{ij}$ and $\sigma_{ij}$ are symmetric, while $\omega_{ij}$ is anti-symmetric.

Taking a derivative of $\theta$ along $\xi$, we find
 \[
    \dot\theta = \nabla_\xi \left( g^{ij}\xi_{ij}\right) = \xi^k g^{ij} \xi_{ijk},
 \]
where for simplicity we have denoted $\nabla_k\xi_{ij} = \xi_{ijk}$. From the definition of the Riemannian tensor, we
have
\[
  \xi_{ijk} - \xi_{ikj} = -R_{jkim} \xi^m,
 \]
hence we get
\[
  \dot\theta = g^{ij} \xi^k\xi_{ikj} - R_{km}\xi^k\xi^m,
\]
where $R_{km}=g^{ji} R_{jkim}$ are the component of the Ricci tensor. Also
\[
  g^{ij} \xi^k\xi_{ikj} = g^{ij} \nabla_j (\xi^k\xi_{ik}) - g^{ij} \xi^k{}_j\xi_{ik} = - \xi^{ki}\xi_{ik}.
\]
Substituting the decomposition of $\xi$ and using the orthogonality relations, we obtain
\[
  -\xi^{ki}\xi_{ik} = -\frac19 \theta^2 h^{ki} h_{ik} - \sigma^{ki}\sigma_{ik} + \omega_{ik}\omega_{ik} = -\frac13
 \theta^2 - \sigma^2 + \omega^2.
\]
Substituting back into the equation for $\dot\theta$ we obtain~\eqref{raychaudhuri}.

Consider now, the Einstein equations $R_{\mu\nu}-\frac12 R g_{\mu\nu}=8\pi G T_{\mu\nu}$. Taking the trace
we get $R=-8\pi GT$, hence substituting back into the Einstein equations we obtain
$R_{\mu\nu}=8\pi G(T_{\mu\nu}-\frac12 T g_{\mu\nu})$  and therefore 
$R_{\mu\nu} U^\mu U^\nu = 8 \pi G(T_{\mu\nu}-\frac12 Tg_{\mu\nu})U^\mu U^\nu$.        
Most known physical matter fields satisfy the Strong Energy Condition (SEC), which states that for all time-like 
vectors $U$, the inequality $T_{\mu\nu}U^\mu U^\nu\geq \frac12 T g_{\mu\nu} U^\mu U^\nu$ holds. It 
follows, when the SEC holds, the term  $R_{\mu\nu}U^\mu U^\nu$ is always positive. Furthermore, note that the shear and 
the rotation are spatial vectors and consequently  $\sigma_{\mu\nu}\sigma^{\mu\nu}\geq 0$, and  
$\omega_{\mu\nu}\omega^{\mu\nu}\geq 0$. 
As mentioned above, $\omega_{\mu\nu}$ is zero if and only if the congruence is
hypersurface-orthogonal. If that is satisfied the Raychaudhuri equation simplifies to the form
\begin{equation} \label{ray-simp}
 \frac{d\theta}{d\tau} + \frac13 \theta^2  + \sigma^2    =  -R_{\mu\nu} U^\mu U^\nu.
\end{equation}
In order for the LHS  to be negative it must fulfill  the condition  $d\theta/d\tau< - \frac13 \theta^2$  which finally 
leads to the inequality:
\begin{equation}
 \frac{1}{\theta(\tau)} \geq \frac{1}{\theta_0} + \frac13 \tau
\end{equation}

One concludes that any initially converging hypersurface-orthogonal congruence must continue to converge and 
within the finite proper time    $\tau \leq - 3\theta_0^{-1}$   will lead to crossing of geodesics (a caustic) which 
means that matter obeying the SEC cannot cause geodesic deviation but will increase the rate of convergence in 
accordance with the fact that the SEC causes gravitation to be attractive~\cite{carroll2004spacetime}.
The aim to define the entropy by the average convergence/divergence of the geodesics in a congruence  will be tantamount to establish that the SEC will cause initially decreasing entropy to continue to decrease.

The Raychaudhuri equation for the expansion
is a first-order nonlinear Riccati     equation and hence of the same type as equation (4) for which the solution, Equation~\eqref{curvature},   has the same form as the Jacobi equation.
     
If we set $\theta = 3 F ' / F$  the Raychaudhuri equation is transformed to
\begin{equation} \label{ray-ricatti}
 \frac{d^2F}{d\tau^2} + \frac13 \left( R_{\mu\nu} U^\mu U^\nu + 
 \sigma^2 - \omega^2 \right) F = 0,
\end{equation}
which is a harmonic oscillator equation. As pointed out above, $\theta$
may be identified with the derivative of the entropy, so that
according to (2) for the entropy $S=\ln F$ here, $F$ may be identified with an effective geodesic
deviation. 

We recently proved~\cite{strauss2014} that the geodesic deviation 
equation of Jacobi is 
essentially equivalent to that of a harmonic oscillator. The expansion $\theta$  is the rate of growth of the 
cross-sectional area 
orthogonal to the bundle of  geodesics. Increase/decrease of this area is the same as the divergence/convergence of the 
geodesics. The average growth of the cross-sectional area is compatible with the \emph{average  geodesic deviation}.

Kar and Sengupta have shown~\cite{kar2007} that the condition for  geodesic convergence is the existence of zeroes in F 
at finite 
values of the affine parameter, and they argue that convergence occurs if
\begin{equation}
 R_{\mu\nu} U^\mu U^\nu + \sigma^2 - \omega^2 \geq 0.
\end{equation}
i.e.\ the shear 
accelerates convergence and the rotation obstructs convergence. 

\section{Comments and Conclusion}

\begin{figure}
\includegraphics[width=8.5cm]{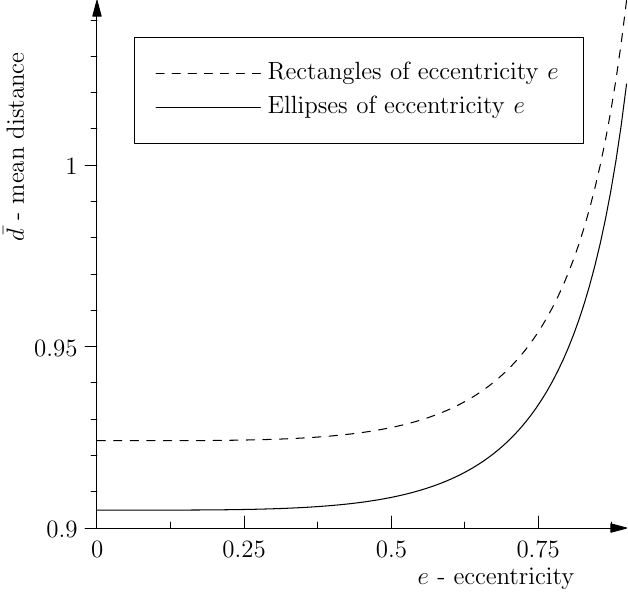}
 \caption{Mean distance between pairs of points}  \label{mean_distance}
\end{figure}

Since shear transforms circles to ellipses, we compared the mean distance between uniformly distributed pairs of independent points inside a circle to that inside an ellipse of the same area, and found that it is smaller in the circle. The mean $\bar d$ of the distance $d$ between pairs of points in a 
planar region $\Omega$ of area $\pi$ can be computed by:
\begin{equation} \label{md}
  \bar d = \frac1{\pi^2} \int_\Omega \int_\Omega |x-y|\,  dA_x dA_y.
\end{equation}
The result is graphed against the eccentricity $e$ in Figure~\ref{mean_distance}. For 
comparison, 
we also computed the same quantity for rectangles of 'eccentricity' $e$ and area $\pi$, where by similarity with the 
definition for an ellipse, we defined the eccentricity of a rectangle with sides $a\geq b$ as $e=\sqrt{1-b^2/a^2}$.
In fact, the mean distance between pairs of points inside any plane domain of area $\pi$ is smallest for a 
circle, i.e.\ the circle is the unique minimizer of~\eqref{md} among all planar regions of area $\pi$~\cite{groemer}. This might have important implications.

The evolution from an infinitesimal circular cross-section orthogonal to the flow 
lines to an elliptical one of same area is brought about by shear. Moving the cross-section along the 
flow does not change the number of geodesics. However, due to increase in the mean-distance between the geodesics
when transforming from a circular to an elliptical cross section,  there is a diverging tendency of the 
geodesics moving  along the flow. That implies, according to our proposed definition of entropy as  
the mean distance between geodesics in a bundle, that the evolution in the presence of shear exhibits an increase of
entropy.

\begin{thebibliography}{9}
\providecommand{\natexlab}[1]{#1}
\providecommand{\url}[1]{\texttt{#1}}
\expandafter\ifx\csname urlstyle\endcsname\relax
  \providecommand{\doi}[1]{doi: #1}\else
  \providecommand{\doi}{doi: \begingroup \urlstyle{rm}\Url}\fi

\bibitem[Benettin et~al.(1976)Benettin, Galgani, and Strelcyn]{benettin1976}
G.~Benettin, L.~Galgani, and J.M. Strelcyn.
\newblock {Kolmogorov entropy and numerical experiments}.
\newblock \emph{Phys. Rev. A}, 14:\penalty0 2338--2345, 1976.
\newblock \doi{10.1103/PhysRevA.14.2338}.
\newblock URL \url{http://link.aps.org/doi/10.1103/PhysRevA.14.2338}.

\bibitem[Carroll(2004)]{carroll2004spacetime}
S.~Carroll.
\newblock \emph{{Spacetime and geometry: an introduction to general
  relativity}}.
\newblock Addison Wesley, San Francisco, 2004.
\newblock ISBN 0805387323.

\bibitem[Casartelli et~al.(1976)Casartelli, Diana, Galgani, and
  Scotti]{casartelli1976}
M.~Casartelli, E.~Diana, L.~Galgani, and A.~Scotti.
\newblock {Numerical computations on a stochastic parameter related to the
  Kolmogorov entropy}.
\newblock \emph{Phys. Rev. A}, 13:\penalty0 1921--1925, 1976.
\newblock \doi{10.1103/PhysRevA.13.1921}.
\newblock URL \url{http://link.aps.org/doi/10.1103/PhysRevA.13.1921}.

\bibitem[{Evangelidis} and {Neethling}(1983)]{Evangelidis1983}
E.~A. {Evangelidis} and J.~D. {Neethling}.
\newblock {On the existence of an entropy-like quantity}.
\newblock \emph{Astrophysics and Space Science}, 96:\penalty0 227--229, 1983.
\newblock \doi{10.1007/BF00661958}.

\bibitem[Gutzwiller(1990)]{gutzwiller2014chaos}
M.~Gutzwiller.
\newblock \emph{{Chaos in classical and quantum mechanics}}.
\newblock Springer, 1990.
\newblock ISBN 978-1-4612-6970-0.

\bibitem[Horwitz et~al.(2007)Horwitz, Zion, Lewkowicz, Schiffer, and
  Levitan]{horwitz2007}
L.~Horwitz, Y.~Zion, M.~Lewkowicz, M.~Schiffer, and J.~Levitan.
\newblock {Geometry of Hamiltonian Chaos}.
\newblock \emph{Phys. Rev. Lett.}, 98:\penalty0 234301, 2007.
\newblock \doi{10.1103/PhysRevLett.98.234301}.
\newblock URL \url{http://link.aps.org/doi/10.1103/PhysRevLett.98.234301}.

\bibitem[Kar and Sengupta(2007)]{kar2007}
S.~Kar and S.~Sengupta.
\newblock {The Raychaudhuri equations: A brief review}.
\newblock \emph{Pramana}, 69\penalty0 (1):\penalty0 49--76, 2007.
\newblock ISSN 0304-4289.
\newblock \doi{10.1007/s12043-007-0110-9}.
\newblock URL \url{http://dx.doi.org/10.1007/s12043-007-0110-9}.

\bibitem[Strauss et~al.(2015)Strauss, P.~Horwitz, Levitan, and
  Yahalom]{strauss2014}
Y.~Strauss, L.~P.~Horwitz, J.~Levitan, and A.~Yahalom.
\newblock {{Q}uantum {F}ield {T}heory of {C}lassically {U}nstable {H}amiltonian
  {D}ynamics}.
\newblock \emph{to appear in J. Math. Phys.}, 2015.
\newblock arXiv:1407.5263v1.

\bibitem[Tabor(1989)]{tabor1989chaos}
M.~Tabor.
\newblock \emph{{Chaos and integrability in nonlinear dynamics : an
  introduction}}.
\newblock Wiley, New York, 1989.
\newblock ISBN 0471827282.

\end{thebibliography}


\begin{bibdiv}
\begin{biblist}

\bib{benettin1976}{article}{
      author={Benettin, G.},
      author={Galgani, L.},
      author={Strelcyn, J.M.},
       title={{Kolmogorov entropy and numerical experiments}},
        date={1976},
     journal={Phys. Rev. A},
      volume={14},
       pages={2338\ndash 2345},
         url={http://link.aps.org/doi/10.1103/PhysRevA.14.2338},
}

\bib{carroll2004spacetime}{book}{
      author={Carroll, S.},
       title={{Spacetime and geometry: an introduction to general relativity}},
   publisher={Addison Wesley},
     address={San Francisco},
        date={2004},
        ISBN={0805387323},
}

\bib{casartelli1976}{article}{
      author={Casartelli, M.},
      author={Diana, E.},
      author={Galgani, L.},
      author={Scotti, A.},
       title={{Numerical computations on a stochastic parameter related to the
  Kolmogorov entropy}},
        date={1976},
     journal={Phys. Rev. A},
      volume={13},
       pages={1921\ndash 1925},
         url={http://link.aps.org/doi/10.1103/PhysRevA.13.1921},
}

\bib{Evangelidis1983}{article}{
      author={{Evangelidis}, E.~A.},
      author={{Neethling}, J.~D.},
       title={{On the existence of an entropy-like quantity}},
        date={1983},
     journal={Astrophysics and Space Science},
      volume={96},
       pages={227\ndash 229},
}

\bib{groemer}{article}{
      author={Groemer, H.},
       title={{On the average size of polytopes in a convex set}},
        date={1982},
     journal={Geometriae Dedicata},
      volume={13},
       pages={47\ndash 62},
}

\bib{gutzwiller2014chaos}{book}{
      author={Gutzwiller, M.},
       title={{Chaos in classical and quantum mechanics}},
   publisher={Springer},
        date={1990},
        ISBN={978-1-4612-6970-0},
}

\bib{horwitz2007}{article}{
      author={Horwitz, L.},
      author={Zion, Y.},
      author={Lewkowicz, M.},
      author={Schiffer, M.},
      author={Levitan, J.},
       title={{Geometry of Hamiltonian Chaos}},
        date={2007},
     journal={Phys. Rev. Lett.},
      volume={98},
       pages={234301},
         url={http://link.aps.org/doi/10.1103/PhysRevLett.98.234301},
}

\bib{kar2007}{article}{
      author={Kar, S.},
      author={Sengupta, S.},
       title={{The Raychaudhuri equations: A brief review}},
    language={English},
        date={2007},
        ISSN={0304-4289},
     journal={Pramana},
      volume={69},
      number={1},
       pages={49\ndash 76},
         url={http://dx.doi.org/10.1007/s12043-007-0110-9},
}

\bib{strauss2014}{article}{
      author={Strauss, Y.},
      author={P.~Horwitz, L.},
      author={Levitan, J.},
      author={Yahalom, A.},
       title={{{Q}uantum {F}ield {T}heory of {C}lassically {U}nstable
  {H}amiltonian {D}ynamics}},
        date={2015},
     journal={to appear in J. Math. Phys.},
      eprint={1407.5263},
        note={arXiv:1407.5263v1},
}

\bib{tabor1989chaos}{book}{
      author={Tabor, M.},
       title={{Chaos and integrability in nonlinear dynamics : an
  introduction}},
   publisher={Wiley},
     address={New York},
        date={1989},
        ISBN={0471827282},
}

\end{biblist}
\end{bibdiv}

\end{document}